# Skyrmion stabilization at the domain morphology transition in ferromagnet/heavy metal heterostructures with low exchange stiffness

J.A. Brock and E.E. Fullerton

*Center for Memory and Recording Research, University of California San Diego, La Jolla, CA, 92093 USA*

**Abstract:**
We report the experimental observation of micron-scale magnetic skyrmions at room temperature in several Pt/Co-based thin film heterostructures designed to possess a low exchange stiffness, perpendicular magnetic anisotropy, and a modest interfacial Dzyaloshinskii-Moriya interaction (iDMI). We find both experimentally and by micromagnetic and analytic modeling that the combined action of low exchange stiffness and modest iDMI eliminates the energetic penalty associated with forming domain walls in thin film heterostructures. When the domain wall energy density approaches negative values, the remanent domain morphology transitions from a uniform state to a labyrinthian stripe phase. A low exchange stiffness, indicated by a reduction in the Curie temperature below 400 K, is achieved in Pt/Co, Pt/Co/Ni, and Pt/Co/Ni/Re structures by reducing the Co thickness to the ultrathin limit (< 0.3 nm). A similar effect occurs in thicker Pt/Co/Ni$_x$Cu$_{1-x}$ structures when the Ni layer is alloyed with Cu. At this transition in domain morphology, skyrmion phases are stabilized when a small (< 1 mT) perpendicular magnetic field is applied and current-induced skyrmion motion including the skyrmion Hall effect is observed. The temperature and thickness-induced morphological phase transitions observed are similar to the well-studied spin reorientation transition that occurs in the ultrathin limit, but we find that the underlying energy balances are substantially modified by the presence of an iDMI.

**Introduction:**

In recent years, there has been an intense study of the magnetic textures known as skyrmions, given that their quantized, winding-like spin structure leads to novel behaviors that are attractive from both fundamental and applied viewpoints.[1,2] Particularly, the unique spin topology of skyrmions lends them a degree of topological stability and offers avenues towards their efficient motion[3] – making them potentially relevant in developing next-generation data storage and novel computation schemes.[4] Previous work has shown that while it is possible to stabilize skyrmion phases through an intricate balance of the ferromagnetic exchange, dipolar, and anisotropy energies of a material system[5], the Dzyaloshinskii-Moriya interaction (DMI)[6,7] – an antisymmetric exchange interaction that favors the orthogonal canting of neighboring magnetic spins – often plays a crucial role. For example, the DMI associated with the non-centrosymmetric $B_{20}$-type crystal structure was central to the first experimental observation of a skyrmion phase.[8]

The interfacial DMI (iDMI) that develops at the interface between thin ferromagnetic (FM) layers and heavy metals (HM) with significant spin-orbit coupling[9] also assists with the formation of skyrmion phases. There have been numerous reports on multi-repetition FM/HM heterostructures with perpendicular magnetic anisotropy (PMA), where a large iDMI assists with the formation of skyrmion phases, and the multiple repetitions of the unit structure increases the total volume of magnetic material, leading to enhanced skyrmion stability.[10,11,12,13,14] Additionally, there have been several reports of skyrmion phases in single magnetic layer HM/FM/HM systems with a modest iDMI, where the FM layer thickness was chosen to be large enough that the system is close to experiencing a spin reorientation transition (SRT)[15,16,17,18], and consequentially has a low effective PMA ($K_{eff}$= $K_u$ - $\frac{\mu_0 M_s^2}{2}$, where $K_u$ is the intrinsic anisotropy constant and $\frac{\mu_0 M_s^2}{2}$ reflects the shape anisotropy of a thin film).[19,20,21,22,23] The domain wall energy density $\sigma$ in systems with iDMI is often approximated as $4\sqrt{A_{ex}K_{eff}} - \pi|D|$ where $A_{ex}$ and $|D|$ are the exchange stiffness and iDMI energy density, respectively.[24] Thus, by lowering $K_{eff}$ while maintaining $|D|$ or vice versa, it is possible to lower $\sigma$, reducing the energy barrier to forming multidomain states, including the skyrmion phase.[25,26]

In this work, we explore the approach of reducing $A_{ex}$ to lower the energy of forming domain walls and facilitate transitions to multidomain and skyrmion phases in ultrathin films while maintaining a reasonable $K_{eff}$ and modest $D$. Given that the domain wall width is typically estimated as $\lambda = \sqrt{A_{ex}/K_{eff}}$, lowering $A_{ex}$ while maintaining significant $K_{eff}$ to obtain skyrmion phases has the added benefit of maintaining a lower domain wall width than would be expected from lowering $K_{eff}$. We report the observation of skyrmion phases at room temperature in several FM/HM heterostructures, including Pt/Co, Pt/Co/NiCu, Pt/Co/Ni, and Pt/Co/Ni/Re, with either one or two repeats of the unit structures listed. While these films are diverse in composition, they share several commonalities, including a modest iDMI arising from the asymmetry of the FM/HM interfaces, significant PMA, and a low exchange stiffness. For the Pt/Co, Pt/Co/Ni, and Pt/Co/Ni/Re samples, the exchange stiffness is lowered through the use of ultrathin (< 0.3 nm) Co layers.[27,28] Similar properties were obtained in a thicker Pt/Co/NiCu sample by increasing the proportionality of Cu to Ni in the alloy layer.[29,30] Concomitant to these changes in the static magnetic properties, we find that as the calculated domain wall energy density diminishes towards negative values, the domain morphology transitions dramatically, permitting the stabilization of labyrinthian stripe domains at remanence and field-induced magnetic skyrmion phases within a narrow window of environmental conditions (*i.e.*, temperature or applied magnetic field). By exploiting the spin-orbit torques generated when passing an electrical current through the sample, skyrmion motion including the skyrmion Hall effect is observed. Using micromagnetic and analytic modeling, we demonstrate the strong relationship between the exchange stiffness and the equilibrium domain morphology in thin films with modest iDMI and significant PMA. Additionally, we show how the associated changes in the energy landscape allow for the stabilization of field-induced skyrmion phases within a framework of understanding like that of previous material approaches[19,20,21,22,23], but through control of an unconventional parameter – the exchange stiffness.

**Experimental Methods:**

Samples of the structure [Co ($t_{Co}$)/ Pt (1)]$_{1,2}$, [Co ($t_{Co}$)/ Ni (0.3)/ Re (0.5)/ Pt (0.5)]$_2$, and [Co ($t_{Co}$)/ Ni (0.3)/ Pt (1)]$_2$, and [Co(0.4)/Ni$_{40}$Cu$_{60}$(0.9)]$_1$ (thicknesses in nm) are used in this study. In the text, these samples will be referred to as the [Pt/Co]$_{1,2}$, [Pt/Co/Ni/Re]$_2$, [Pt/Co/Ni]$_2$, and [Pt/Co/NiCu]$_1$ samples, respectively. A Ta (2)/ Pt (5) seeding layer and Ta (3) capping layer were used for all samples. The samples were grown using dc magnetron sputtering at ambient temperature onto Si substrates with a 300 nm-thick thermal oxide coating. A 3 mTorr partial pressure of Ar and a sputtering power of 50W were used. For some samples, the Co layers were grown as wedge-type structures using a slit placed on top of the Co sputtering source – allowing for a 14 % thickness gradient to be achieved during a single deposition over a 4 cm-long substrate. Some samples were patterned into 100-μm-wide wires using conventional metal lift-off UV photolithography. Temperature-dependent magnetometry was performed in the out-of-plane (OOP) and in-plane (IP) geometries using vibrating sample magnetometry (VSM). When calculating volumetric parameters from the magnetometry data, we assume that any Pt within 0.2 nm of an interface with a FM layer has become magnetically polarized.[31] Magneto-optic Kerr effect (MOKE) images and hysteresis loops were obtained using an Evico Magnetics microscope, in which perpendicular magnetic fields were applied using an air coil. For some MOKE measurements, the sample temperature was varied using a Peltier chip. For the MOKE images presented within, images of the sample when the magnetization was saturated have been subtracted as a background. Micromagnetic modeling was performed using the MuMax3 solver[32], employing a 5 μm x 5 μm x 1.5 nm geometry discretized into 2 nm x 2 nm x 1.5 nm cells. Periodic boundary conditions were used to account for ten repetitions of the simulation geometry within the film plane.

**Experimental Results:**

We begin by discussing experimental results for the [Pt/Co]$_2$ samples. While the Pt/Co/Pt structure is nominally symmetric, previous work has shown that structural asymmetry between the Pt-Co and Co-Pt interfaces can lead to the development of PMA and a modest iDMI.[33,34] From measurements of the domain expansion velocity as a function of in-plane field for a [Pt/Co(0.35 nm)]$_2$ sample, we observe asymmetries that are typically associated with the "left-handed" iDMI thought to originate from the lower Pt-Co interface[34] and estimate (using techniques described elsewhere[24,35]) a $|D|$ of ~ 0.14 mJ/m$^2$ (see Supp. Note 1 and Supp. Fig. 1[36]). In agreement with previous studies, the saturation magnetization as a function of temperature $M_s(T)$ plots provided in Fig. 1(a) show a strong dependence on $t_{Co}$, and that $T_C$ decreases as $t_{Co}$ is reduced within the ultrathin limit.[27,28] In particular, Fig. 1(a) demonstrates that the sample with $t_{Co}$ = 0.32 nm is thermally stable at room temperature, as $T_C$ is well above 300 K. However, decreasing $t_{Co}$ to 0.24 nm suppresses the room temperature $M_s$, and the $M_s(T)$ curve demonstrates a $T_C$ of ~365 K – indicating a significant lowering of $A_{ex}$. Further decreasing $t_{Co}$ to 0.2 nm suppresses $T_C$ to near room temperature. While it is anticipated that the uniformity of the Co layers may be compromised in the ultrathin limit, previous reports have shown that well-defined FM behavior is observed even with sub-monolayer FM coverage, provided the FM layer is abutted by materials susceptible to proximity-induced magnetism (*e.g.*, Pt, Pd).[27,28]

In Fig. 1(b), we show room temperature polar MOKE hysteresis loops for [Pt/Co]$_2$ samples with different $t_{Co}$. As expected for FM/HM heterostructures with PMA, we find that films with $t_{Co} \geq 0.26$ nm have square hysteresis loops. However, reducing $t_{Co}$ to 0.25 nm or 0.24 nm, sheared polar MOKE loops are observed – suggesting a multidomain state at remanence in these samples. Further reducing $t_{Co}$ to a nominal thickness of 0.23 nm, the polar MOKE loop has zero remanent magnetization and a hard axis-like character; as Fig. 1(a) demonstrated that $T_C$ decreases dramatically in this regime of $t_{Co}$, this may indicate a paramagnetic-like character to this sample. To understand how these changes in the static magnetic properties impact the room temperature domain morphology, we have collected polar MOKE images at zero field [Figs. 1(c-f)]. Before capturing an image, the following field history protocol was employed: First, the sample magnetization was saturated in the negative perpendicular field direction (-$M_z$). The magnetic field was then gradually increased towards each sample's respective positive coercive field ($+H_c$) before removing the magnetic field. For samples with $t_{Co} \geq 0.26$ nm, circular magnetic domains were observed at remanence – as is expected for thin films with PMA. In agreement with the polar MOKE hysteresis loops, a labyrinthian stripe domain pattern with a sporadic density of circular domain textures was observed at remanence when $t_{Co}$ was reduced to 0.24 nm. Further reducing $t_{Co}$ to 0.23 nm, the featureless remanent state may indicate an in-plane magnetic anisotropy, a paramagnetic-like character to this sample, or that the domain morphology is smaller and/or fluctuating faster than our polar MOKE system can resolve. Overall, Figs. 1(c-f) demonstrate that the domain morphology evolves dramatically as a function of thickness in the limit of ultrathin Co.

In Figs. 2(a,b), we provide more detailed, temperature-dependent magnetometry data for the [Pt/Co(0.24 nm)]$_2$ sample. From OOP hysteresis loops collected in the temperature range of 294 K to 302 K [Fig. 2(a)], the extreme temperature sensitivity of this sample's magnetic properties can be seen. For $T \leq 292$ K, square hysteresis loops are observed, as is common for thin films with PMA. As the temperature is increased over the range 294 K $\leq T \leq 302$ K, the hysteresis loop become progressively more sheared – suggesting a multidomain remanent state. For $T \geq 302$ K, the hysteresis loops take on a hard axis-like character (*i.e.*, low remanence, low hysteresis). Thus, the evolution of the OOP hysteresis loops with increasing temperature is qualitatively similar to the evolution with decreasing Co thickness shown in Fig. 1(b). The corresponding IP hysteresis loops, collected in the same temperature range over which the OOP loops exhibited such dramatic changes, are shown in Fig. 2(b). Surprisingly, we find that the IP saturation field is relatively large, giving an anisotropy field (estimated from the saturation field) of $\mu_0 H_K \approx 500$ mT that does not appreciably change with temperature – suggesting that the temperature dependence of $K_{eff}$ is primarily scaled by $M_s$. The magnetometry data collected at $T = 302$ K shown in Figs. 2(a) and 2(b) demonstrate that both the OOP and IP hysteresis loops have zero remanence and little to no hysteresis. This evolution in magnetic response with temperature, including the formation of stripe domains, is similar in many respects to that observed during the spin reorientation transition (SRT) that occurs in the limit of ultrathin FM layers, when the intrinsic PMA equals the shape anisotropy (*i.e.*, when $K_{eff} = 0$).[15,16] However, above the SRT temperature, a finite remanence and low saturation field is expected in the IP loops – behavior that is quite different from our observations. A further comparison of our results to the previously studied SRT is provided in the discussion section.

In Figs. 1(c-f), we show polar MOKE images of the [Pt/Co(0.24 nm)]$_2$ sample as the perpendicular magnetic field strength $\mu_0 H_z$ was quasi-statically decreased towards negative saturation ($-M_z$). As mentioned previously, a labyrinthian domain pattern with a sporadic population of circular features is present in this sample when $\mu_0 H_z = 0$. As in other systems whose static magnetic properties are extremely sensitive to temperature, we observe that the remanent domain pattern experiences a significant degree of thermal fluctuation when observed over several seconds[19], particularly for the [Pt/Co]$_1$ sample (see Supp. Video 1[36]). When a sufficiently strong $\mu_0 H_z$ is applied, some stripe domains collapse into circular features with a characteristic diameter of ~ 3 μm. Further increasing $\mu_0 H_z$ causes all the stripe domains to collapse, forming a disordered ensemble of circular textures. As $\mu_0 H_z$ approaches the negative saturation field, the circular features begin to disappear, and eventually, the film exhibits a uniform magnetization. A video of this evolution in domain morphology with $\mu_0 H_z$ is provided in Supp. Video 2.[36] While this field-induced stripe-to-skyrmion transition can be observed at room temperature in the [Pt/Co(0.24 nm)]$_2$ sample, we note that by heating the [Pt/Co(0.26 nm)]$_2$ sample (*i.e.*, the sample whose room temperature domain morphology is shown in Fig. 1(c)) slightly, similar behavior is observed (Supp. Fig 2[36]).

To confirm that these field-induced circular features are magnetic skyrmions with chiral domain walls, their response to applied electrical currents was determined. As the samples were seeded and capped with HM materials thought to possess spin-Hall angles of opposing sign (Pt and Ta, respectively)[37,38], a net spin-orbit torque is expected to act on the FM layers when a current is passed through the structure; as such, the skyrmions should move collinear to an applied electrical current (not considering any other dynamic effects). Furthermore, if the features possess the quantized topology characteristic to skyrmions, they should also experience a Magnus force when driven, leading to a deflection transverse to the applied current axis – a phenomenon known as the skyrmion Hall effect.[39,40,41] As demonstrated in Fig. 3 (and Supp. Videos 3 and 4[36]), the circular features stabilized by $\mu_0 H_z$ in the [Pt/Co(0.24 nm)]$_2$ sample move in the direction of the conventional current density $J$; knowing the spin-Hall angles of the Pt seeding and Ta capping layers, this response is consistent with spin-orbit torques acting on the "left-handed" chirality observed in other Pt/Co-based systems.[19,42,43,44,45,46] As shown in Supp. Video 4, we also observed a dramatic increase in the number of skyrmions present in the wire after an electrical current has been applied – a behavior that may be attributed to the previously reported mechanisms by which skyrmions can be generated from stripe domains subject to spin-orbit torques and sporadic pinning sites.[47] The skyrmion Hall effect is discerned using a technique similar to that employed in previous works[19], whereby skyrmions were ejected from a 100 μm-wide wire (where the skyrmions move with a speed too fast for our system to image) into a 3 mm-wide contact pad where the skyrmions slow, accentuating any transverse motion. From Fig. 3 (and Supp. Video 5[36]), it can be seen that the features exhibit a transverse deflection relative to the applied current axis and that the direction of this deflection reverses when changing between +$M_z$ and -$M_z$ polarity domains, as would be expected from the skyrmion Hall effect.[19,41]

Considering these findings in [Pt/Co] samples in the limit of ultrathin $t_{Co}$, we now examine how modifying the HM and FM layers to increase the structural inversion asymmetry (and hence, the total iDMI) impacts the morphological phases accessible in the ultrathin limit. We highlight two other material systems: [Pt/Co(0.24 nm)/Ni/Re]$_2$ and [Pt/Co(0.24 nm)/Ni]$_2$ (thicknesses in nm) – all featuring interfaces (Pt-Co, Ni-Re, and Ni-Pt) thought to contribute to an additive iDMI acting on the FM layers due to their compositional asymmetry (*i.e.*, in addition to an asymmetry in interfacial quality, as in the [Pt/Co] samples).[48,49] While the Co layers of the two structures mentioned above were grown as wedges, we indicate the thickest approximate $t_{Co}$ at which a $\mu_0 H_z$-induced skyrmion phase was observed at room temperature. The remanent state of the [Pt/Co/Ni/Re]$_2$ sample shown in Fig. 4(a) succinctly demonstrates the compositional sensitivity of the domain morphology in the ultrathin limit, as the stripe domain size dramatically decreases when $t_{Co}$ is varied by as little as ~ 0.05 %. We emphasize that while Fig. 4(a) is reminiscent of previous studies that showed a strong link between film thickness and domain periodicity on account of proximity to a SRT[15,16,17,18], the static characterization does not indicate the presence of the SRT in our FM/HM heterostructure samples near room temperature, given the large in-plane saturation field indicating a significant PMA.

From the higher-magnification polar MOKE images shown in Figs. 4(b, d), it can be seen that the [Pt/Co/Ni/Re]$_2$, and [Pt/Co/Ni]$_2$, samples exhibit a labyrinthian stripe pattern in zero field at room temperature. In line with the [Pt/Co]$_2$ samples, temperature-dependent magnetometry of these samples indicates $T_C$ values below 400 K and OOP hysteresis loops extremely sensitive to variations in temperature in the vicinity of 300 K. Much like the [Pt/Co]$_2$ samples, the stripe domains formed at remanence in the [Pt/Co/Ni/Re]$_2$ and [Pt/Co/Ni]$_2$ samples transform into circular textures when small perpendicular fields are applied. By observing these features' response to spin-orbit torques (using the process previously described for the [Pt/Co]$_2$ samples), it is possible to classify these features as skyrmions. Figures 4(c, e) show polar MOKE images collected at the $\mu_0 H_z$ value where each sample exhibits its densest skyrmion phase.

Thus far, the materials systems discussed have shared the commonalities of an ultrathin (< 0.3 nm) Co layer and a low $A_{ex}$ (inferred from a low $T_C$). However, if a low $A_{ex}$ is responsible for the observed domain morphologies, it should be possible to attain similar results in systems with thicker FM layers. Previous works have shown that alloying Ni with Cu can lower $T_C$ relative to that of elemental Ni and that this trend is linear with respect to Cu concentration.[29] Based on this, we fabricated a Ta(2)/Pt(5)/Co(0.4)/Ni$_{40}$Cu$_{60}$(0.9)/Ta(3) sample (thicknesses in nm) to explore the role of exchange stiffness by optimizing the Cu-Ni ratio. From magnetometry data collected near room temperature [Fig. 5(a)], it can be seen that the [Pt/Co/NiCu]$_1$ sample's OOP hysteresis loops are extremely sensitive to temperature; the square loop present at 285 K becomes more sheared as the temperature is increased to 290 K, and becomes hard axis-like when the temperature is increased to 295 K. These findings are similar to those shown for the [Pt/Co(0.24 nm)]$_2$ sample shown in Fig. 2(a).

Much like the [Pt/Co]$_2$ samples with ultrathin Co layers, the IP hysteresis loops of the [Pt/Co/NiCu]$_1$ sample do not change significantly in the vicinity of room temperature [Fig. 5(b)]; given that the sign of $K_{eff}$ does not change over the relevant temperature range, this does not indicate the presence of a SRT near room temperature. The $M_S(T)$ curve shown in Fig. 5(c) (generated from hysteresis loops collected over a wide range of temperatures) indicates a $T_C$ of approximately 360 K for the [Pt/Co/NiCu]$_1$ sample. From polar MOKE images collected at $T$ = 290 K, it can be seen that the [Pt/Co/NiCu]$_1$ sample exhibits a labyrinthian stripe domain pattern at remanence near room temperature, with a characteristic domain width of ~ 2 μm [Fig. 5(d)]. When a ~ 0.04 mT perpendicular field is applied, the labyrinthian remanent state is transformed into an ensemble of circular textures [Fig. 5(e)] that display the spin-orbit torque induced dynamics emblematic of skyrmions as seen in the previously discussed samples with ultrathin Co layers – indicating that the energetic considerations that give rise to the remanent labyrinthian morphology and field-induced skyrmion phase are not contingent on a reduced dimensionality to the FM layers.

**Discussion:**

Thus far, the temperature dependence of the OOP hysteresis loops, the suppressed $T_C$ values, and the labyrinthian domain morphologies observed at remanence in our samples draw strong parallels to extensive, past studies of the SRT.[15,16,18] Particularly, the thickness sensitivity exhibited in Fig. 4(a) is heavily reminiscent of previous studies of a Cu/Fe/Ni structure that was found to exhibit a SRT when the Fe thickness was reduced to ~ 1.25 ML.[18] While the original understanding of the SRT was developed using systems that do not have iDMI, recent reports have shown that the iDMI can assist in raising the temperature at which the SRT occurs and facilitates a direct transition between the ferromagnetic and paramagnetic states for sufficient $D$.[50] However, despite the many similarities between our HM/FM heterostructures and systems that undergo a SRT, our work is differentiated by the fact that $K_{eff}$ does not approach zero when the labyrinthian stripe phase becomes the remanent state in our samples (whether by adjusting the FM layer thickness or the sample temperature) – a key point that rules out a traditional SRT at the thicknesses and temperatures relevant to our samples. Given that a harbinger of a skyrmion phase in our samples has been the observation of a labyrinthian stripe phase at remanence and the treatment of the SRT, while related, may not be applicable, further exploration is warranted.

To better understand the factors that give rise to skyrmion phases in thin films with significant PMA, modest iDMI, and low $A_{ex}$, we have performed micromagnetic and analytic modeling. For our models, a $M_S$ of 250 kA/m, a first-order uniaxial anisotropy constant $K_u$ of 67.5 kJ/m$^2$ (corresponding to $K_{eff} = K_u - \frac{\mu_0 M_S^2}{2} = 28.25$ kJ/m$^3$ in the thin film geometry), and a film thickness $t_{film}$ of 1.5 nm were used to approximate the [Pt/Co/NiCu]$_1$ sample. While an $A_{ex}$ of ~ 10 pJ/m is often used when modeling Pt/Co and Pt/Co/Ni-based systems, this generally assumes that $T_C$ is closer to the bulk value of Co.[51,52,53] To reflect that fact that $T_C$ of the [Pt/Co/NiCu]$_1$ sample is ~ 75 % lower than that of bulk Co, we use an $A_{ex}$ of 2.5 pJ/m in our calculations. The iDMI energy density $D$ was varied between 0 and 1 mJ/m$^2$. All simulations were performed at $T$= 0 K.

In the thin film limit, it is well known that the relative balance between the demagnetization energy and the energy penalty associated with forming a domain wall (both of which depend on $M_S$, $K$, $A_{ex}$, and $D$) determines whether the film will break into domains in zero field.[54,55,56,57] To understand this energetic competition in the [Pt/Co/NiCu]$_1$ sample (which can generally be extended to describe the samples with ultrathin Co layers), we first performed a simple micromagnetic calculation of the difference in energy density $\Delta \varepsilon$ between the two-domain state with a single domain wall and the uniform magnetic state (*i.e.*, $\Delta \varepsilon = \varepsilon_{\text{two-domain}} - \varepsilon_{\text{uniform}}$) as a function of $D$. This calculation of $\Delta \varepsilon$ allows for a micromagnetic estimation that can be compared to the analytical expression for the domain wall energy density $\sigma = 4\sqrt{A_{ex} K_{eff}} - \pi |D|$. A schematic depiction of the magnetic states considered is provided in Fig. 6(a). The $\varepsilon$ associated with each state was determined after allowing the domain state to relax from the respective initial state to the minimum energy configuration. From the $\Delta\varepsilon(D)$ plot provided in Fig. 6(b), it can be seen that the uniform state has a lower $\varepsilon$ state at lower $D$ – indicating that the energetic penalty associated with breaking into domains/forming a domain wall outweighs the associated minimization in dipolar energy. However, at a relatively modest value of $D$ (~ 0.28 mJ/m$^2$), $\Delta\varepsilon(D)$ becomes negative, suggesting that the energetic penalty to forming a domain wall has been diminished and the energy landscape is more conducive to the formation of multidomain states. If instead, an $A_{ex}$ of 10 pJ/m is used [Fig. 6(b)] – as is often the case when describing Pt/Co-based systems – it can be seen that the root of $\Delta\varepsilon(D)$ is shifted towards $D$ values which are generally only feasible in material systems optimized to have a large iDMI.[49,58] or by reducing the effective anisotropy. Calculations of the domain wall energy density $\sigma = 4\sqrt{A_{ex}K_{eff}} - \pi|D|$ under the same considerations – also shown in Fig. 6(b) – show similar trends as the $\Delta\varepsilon(D)$ simulations. We note that accounting for the domain wall anisotropy energy density in the net $\sigma$ only slightly modifies the curve shown in Fig. 6(b). While these results are most applicable to the [Pt/Co/NiCu]$_1$ sample, we show how these trends in $\Delta\varepsilon$ and $\sigma$ evolve with $A_{ex}$, $K$, and $D$ with greater detail in Supp. Fig. 3[36]. In keeping with our experimental results, as $A_{ex}$ and $D$ are not expected to change dramatically when varying the temperature by several K[59,60], we argue that the reduction in $\Delta\varepsilon/\sigma$ with temperature in our samples is primarily determined by $K_{eff}$, which itself is scaled by $M_s$ (not $\mu_0 H_K$). Indeed, $M_s$ of the [Pt/Co/NiCu]$_1$ sample changes dramatically for small modifications in temperature within the relevant temperature range [Fig. 5(c)]. Thus, while the shift in relative energy of the uniform magnetic state versus a multidomain state is driven by $K_{eff}$, a low $A_{ex}$ and modest $D$ permit this transition without $K_{eff}$ approaching zero (*i.e.*, away from the SRT).

To understand how these energetic balances impact the equilibrium domain size in the [Pt/Co/NiCu]$_1$ sample, we have employed the analytic model of Ref. 61. In Fig. 6(c), we show the equilibrium domain width predicted as a function of $t_{film}$ for the [Pt/Co/NiCu]$_1$ sample (using the same static parameters as those used in the micromagnetic modeling) for several values of $D$. For the lower $D$ values (0 and 0.26 mJ/m$^2$), a sample with $t_{film}$ = 1.5 nm is predicted to exhibit mm-scale domains, which may be considered as a uniform magnetic state. This calculation agrees with the micromagnetic modeling, which suggested that it was energetically disadvantageous for a sample with $t_{film}$ = 1.5 nm to break into domains at lower $D$. For $D$ = 0.28 mJ/m$^2$ (*i.e.*, close to the $D$ value where $\Delta\varepsilon/\sigma$ becomes negative), the equilibrium domain size decreases exponentially with the film thickness in the vicinity of the $t_{film}$ values relevant to the [Pt/Co/NiCu]$_1$ sample. Knowing that $t_{film}$ = 1.5 nm and the experimentally determined domain size is ~ 2 μm, the analytic model indicates that 0.28 mJ/m$^2$ is a reasonable estimation for $D$ in the [Pt/Co/NiCu]$_1$ film.

Taken in concert, the micromagnetic and analytic modeling suggest that a modest $D$ can enable the formation of a labyrinthian stripe phase at remanence in samples with low $A_{ex}$ – behavior that is not typical to thin films with low $M_s$ and appreciable PMA. However, the appearance of a labyrinthian stripe phase at zero field does not necessarily guarantee the presence of a field-induced skyrmion phase. To understand the nature of the field-induced morphological transition between stripe domains and skyrmions, we have performed additional micromagnetic simulations, whereby we compare the $\mu_0 H_z$-dependence of $\varepsilon$ between the labyrinthian stripe and skyrmion phases at $T$ = 0 K.[62] For these calculations, the material parameters were the same as those used when generating Fig. 6(b) and a $D$ = 0.28 mJ/m$^2$. The initial state of the simulation was set to a random magnetization pattern or bubble lattice (for the labyrinthian stripe and skyrmion phases, respectively). At each $\mu_0 H_z$ step in the simulation, the system was relaxed to its minimum total energy configuration. In line with the experimental results, comparing the $\varepsilon(\mu_0 H_z)$ profiles of these two morphological phases [Fig. 6(d)] demonstrates that the skyrmion phase becomes the ground state when $\mu_0 H_z \approx$ 0.09 mT, which is reasonably close to the experimental findings. While an energetic barrier must be overcome to transition between morphological states, finite-temperature atomistic simulations have shown that bringing the system closer to $T_C$ can shallow this barrier.[62] Given that all the samples discussed herein have a $T_C$ less than 400 K, we argue that the energy barrier separating morphological phases may be easily overcome by the thermal energy present near room temperature.[62,63]

We have yet to comment in detail on the magnetic state present when the hysteresis loops indicate PMA [Figs. 1(b), 2(a,b)] but polar MOKE contrast has disappeared and there is little or no hysteresis [Figs. 1(b,f)]. In the simplest scenario, the domain size at such temperatures or FM thicknesses may be too small to resolve with polar MOKE. In the context of the SRT, however, such observations have been interpreted as an indicator of a phase transition, and there has been extensive debate in the literature as to whether this consists of a transition from the ferromagnetic state to a paramagnetic "gap" state with PMA[18,64] or a fluctuating stripe domain phase[65,66,67]. While our samples do not have the low $K_{eff}$ emblematic of the SRT, the static magnetic properties are similar in many regards; thus, further investigations with higher spatiotemporal resolution are needed to fully understand the nature of the temperature/thickness-induced domain morphology transformations observed in our samples. Regardless of the end result of the transformation, the modeling discussed above permits an understanding of the energetic balances responsible for skyrmion stabilization in our samples with low exchange stiffness, appreciable PMA, and modest iDMI.

**Conclusion:**

In summary, we have experimentally characterized the domain morphologies present in a number of thin film heterostructures with perpendicular magnetic anisotropy, modest interfacial Dzyaloshinskii-Moriya interaction energy density, and low exchange stiffness. We find that by lowering the exchange stiffness, the remanent domain morphology transitions from a uniform state to a labyrinthian stripe phase – drawing strong, yet nuanced parallels to previous work on the spin reorientation transition that occurs in the limit of vanishing effective perpendicular magnetic anisotropy. Furthermore, when a small perpendicular magnetic field is applied at this morphological transition, skyrmion phases become stabilized. Spin-orbit torque-induced motion including the skyrmion Hall effect is observed when an electrical current is passed through the samples. Micromagnetic and analytic modeling demonstrates that in the limit of low exchange stiffness, the presence of a moderate interfacial Dzyaloshinskii-Moriya interaction modifies the energetic balance of thin films, allowing for the formation of multidomain remanent states and a shallowing of the energy barrier separating the stripe and skyrmion ground states.

**Acknowledgements:**

The authors gratefully acknowledge support from the National Science Foundation, Division of Materials Research (Award #: 2105400). This work was performed in part at the San Diego Nanotechnology Infrastructure (SDNI) of the University of California – San Diego, a member of the National Nanotechnology Coordinateed Infrastructure, which is supported by the National Science Foundation (Grant #: ECCS-1542148).

**Main Text Figures:**

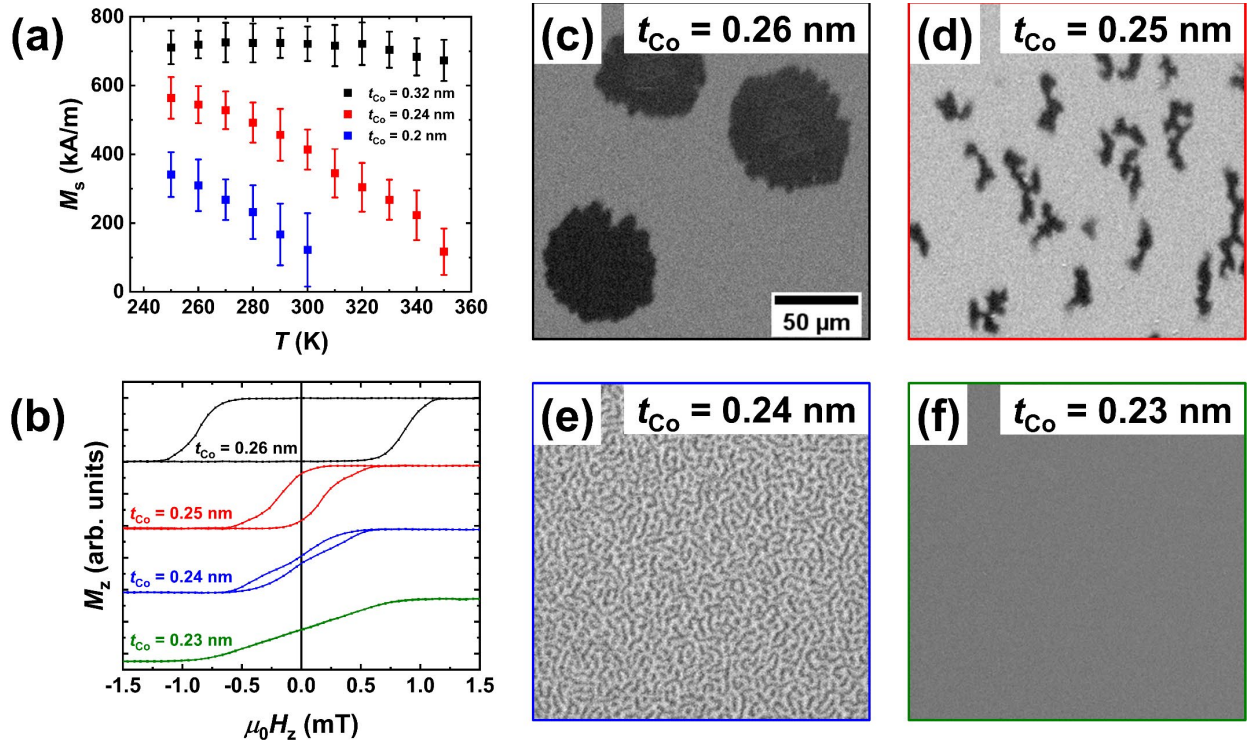

**Figure 1: (a)** Saturation magnetization as a function of temperature for [Pt/Co]$_2$ samples with varying Co thicknesses ($t_{Co}$ = 0.32 nm, 0.24 nm, and 0.2 nm), determined using VSM. **(b)** Polar MOKE loops for [Pt/Co]$_2$ samples with 0.23 nm ≤ $t_{Co}$ ≤ 0.26 nm, collected at room temperature. The MOKE loops have been offset for clarity. Room temperature remanent states of the [Pt/Co]$_2$ samples with **(c)** $t_{Co}$ = 0.26 nm, **(d)** $t_{Co}$ = 0.25 nm, **(e)** $t_{Co}$ = 0.24 nm, and **(f)** $t_{Co}$ = 0.23 nm. The remanent states were attained using the field cycling process discussed in the main text.

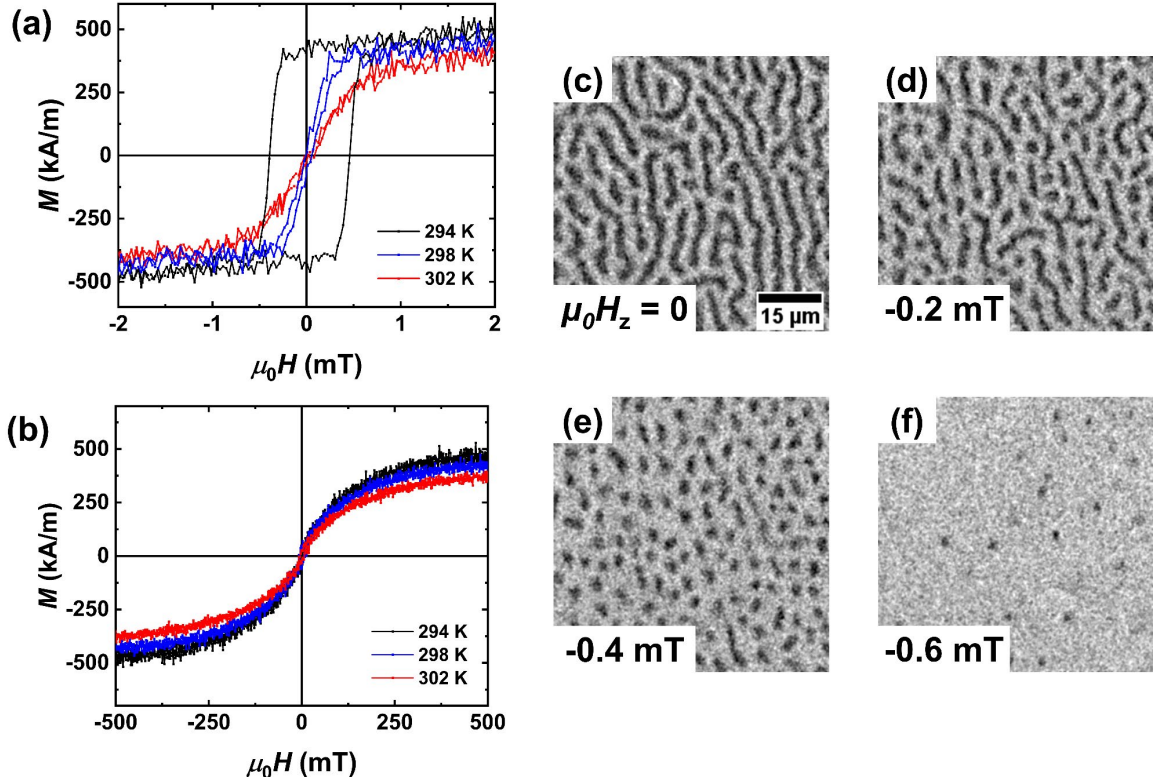

**Figure 2:** For the [Pt/Co(0.24 nm)]$_2$ sample: **(a)** Out-of-plane and **(b)** in-plane magnetometry data collected in the temperature range of 294 K ≤ $T$ ≤ 302 K. **(c-f)** Room temperature polar MOKE images collected as the perpendicular magnetic field strength $\mu_0 H_z$ was quasi-statically stepped from $\mu_0 H_z = 0$ towards negative saturation.

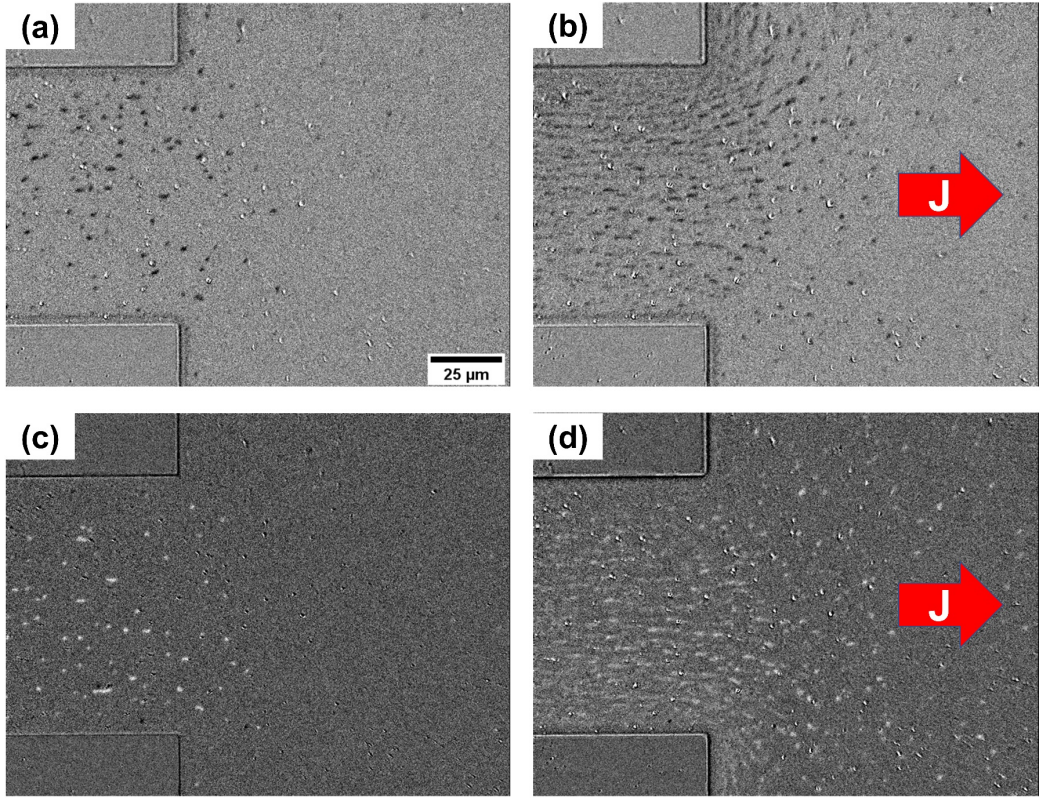

**Figure 3:** **(a)** and **(b)** +$M_z$ skyrmions stabilized in a perpendicular magnetic field of – 0.7 mT in a [Pt/Co(0.24 nm)]$_2$ device photolithographically patterned into a 100 µm-wide wire with 3 mm-wide contact pads **(a)** before and **(b)** while an electrical current density of $J$ = 5.5 x 10$^9$ A/m$^2$ was applied. **(c)** and **(d)** were collected under the same protocol but with $\mu_0 H_z$ = + 0.7 mT to obtain -$M_z$ skyrmions. The images shown in **(b)** and **(d)** represent the average over images collected at 16 Hz while the current was applied for 1 s. The behavior demonstrated in Fig. 3 is provided as a video in Suppl. Video 4.[36]

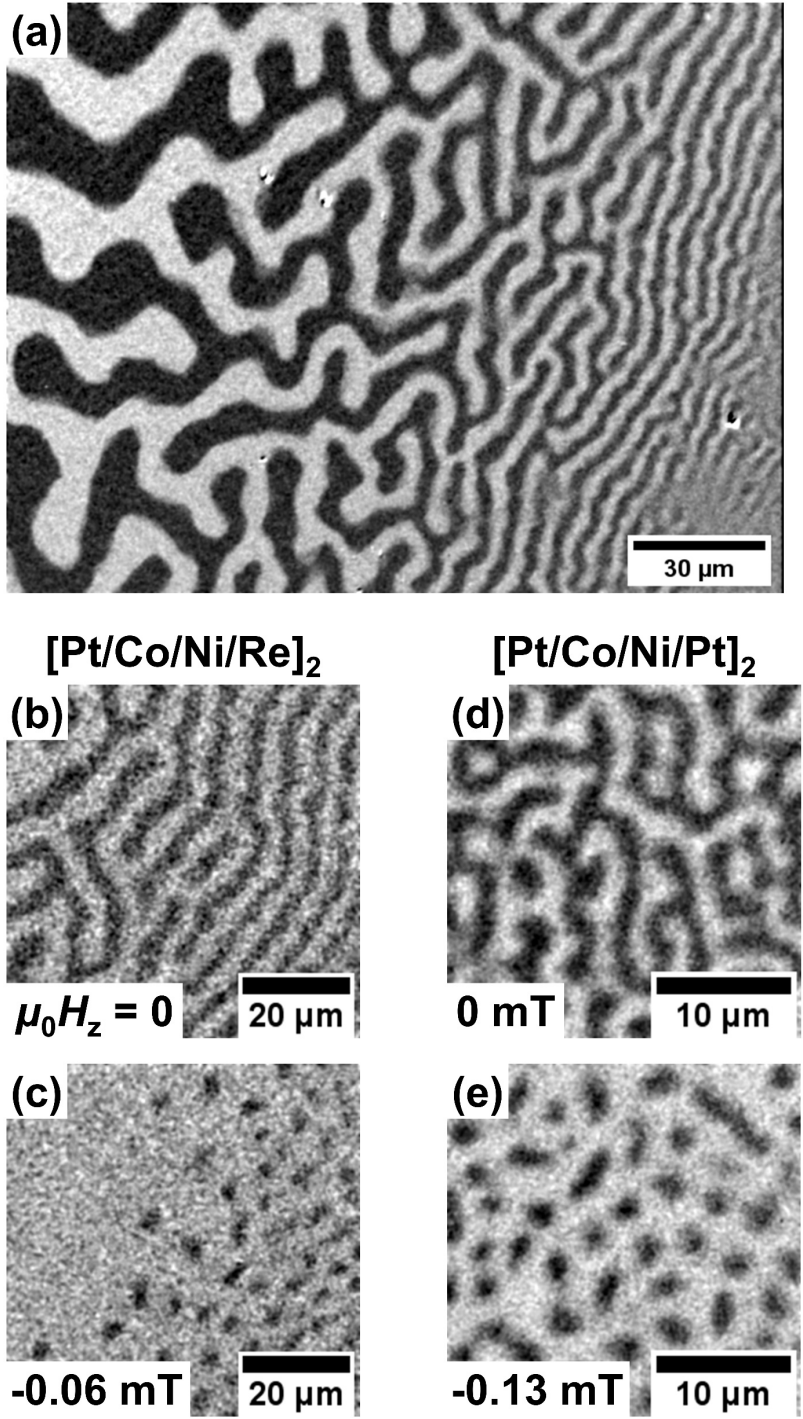

**Figure 4: (a)** Remanent domain morphology of the [Pt/Co/Ni/Re]$_2$ sample, where the Co layers were grown such that $t_{Co}$ decreases by ~ 0.05 % when moving from left to right across the field of view. The center of the field of view corresponds to an approximate $t_{Co}$ of 0.23 nm. Remanent state and densest $\mu_0H_z$-induced skyrmion phases observed in the **(b,c)** [Pt/Co/Ni/Re]$_2$ and **(d,e)** [Pt/Co/Ni]$_2$ samples.

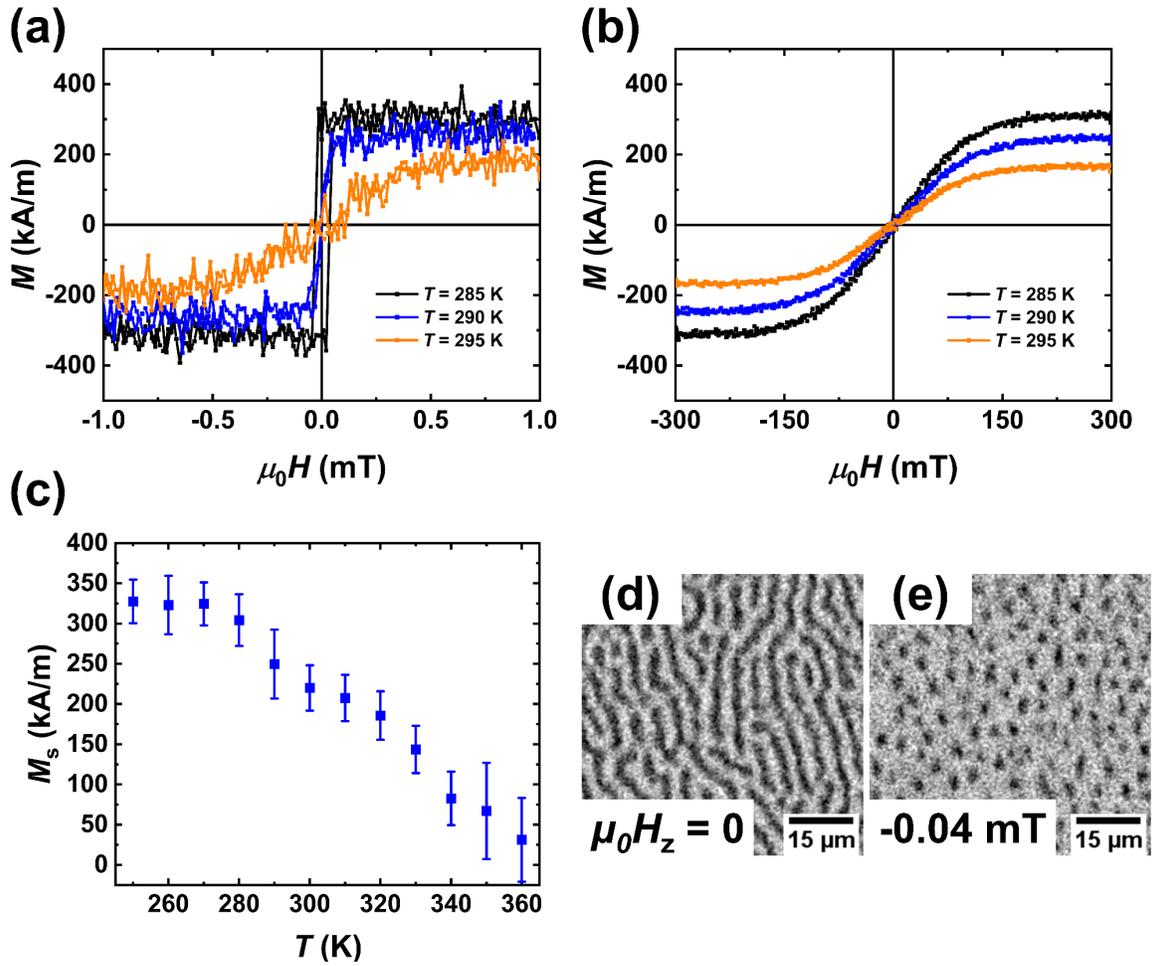

**Figure 5:** For the [Pt/Co/NiCu]$_1$ sample: **(a)** out-of-plane and **(b)** in-plane magnetometry data collected in the temperature range of 285 K ≤ $T$ ≤ 295 K. **(c)** Saturation magnetization $M_s$ over the temperature range 250 K ≤ $T$ ≤ 360 K. Polar MOKE images collected at $T$ = 290 K in perpendicular magnetic fields of 0 and -0.04 mT are shown in **(d)** and **(e)**, respectively.

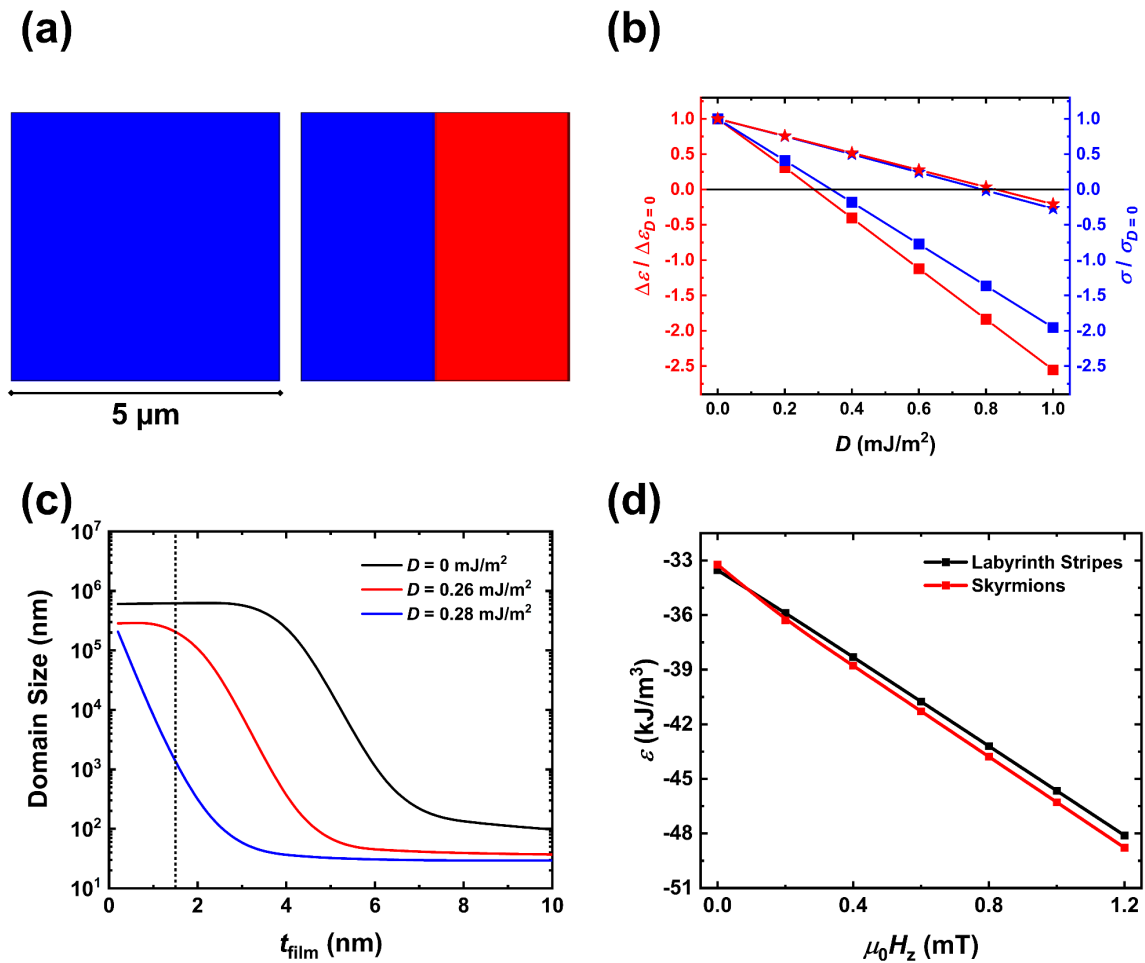

**Figure 6:** Modeling results for the [Pt/Co/NiCu]$_1$ sample: **(a)** Schematic depiction of the uniform (left) and two-domain (right) states used in the micromagnetic simulations. Blue (red) regions correspond to $+M_z$ ($-M_z$) magnetic orientations. **(b)** Micromagnetic calculations of the difference in energy density $\Delta\varepsilon$ between the uniform and two-domain states (red) and the domain wall energy density $\sigma$ (blue) as a function of iDMI energy density $D$. For each initial state, $\Delta\varepsilon/\sigma$ has been normalized to the value when $D = 0$. Data is shown for $A_{ex} = 2.5$ pJ/m (squares) and $A_{ex} = 10$ pJ/m (stars). **(c)** Analytic modeling of the expected domain size as a function of the magnetic film thickness $t_{film}$ for selected $D$ values. The dashed line indicates the magnetic layer thickness of the [Pt/Co/NiCu]$_1$ sample used in our experiments. **(d)** $\varepsilon$ as a function of $\mu_0 H_z$ for the labyrinthian stripe and skyrmion phases. The material parameters used in the modeling are stated in the Discussion portion of the main text.

**Supplemental Information:**

**Supplemental Note 1:**

To estimate the iDMI energy density $D$ of the [Pt/Co]$_2$-type samples, we have measured the in-plane field-dependendence of the velocity $v$ of ↑ to ↓ and ↓ to ↑ domain walls (*i.e.*, the right and left sides (respectively) of the domains shown in Supp. Figs. 1(a,b)) as a pulsed perpendicular magnetic field was applied. As the iDMI promotes chiral Neel-type domain walls (schematically shown in Supp. Fig. 1(b)), an in-plane magnetic field is expected to break the energy symmetry of ↑ to ↓ and ↓ to ↑ domain walls, leading to an asymmetry in the velocity of expansion. As shown in Supp. Fig. 1(c), ↑ to ↓ domains exhibit a minimum in velocity when an in-plane magnetic field $\mu_0 H_x$ is applied in the $+\hat{x}$-direction indicated in Supp. Fig. 1(b). By determining the $\mu_0 H_x$ at which this minimum in velocity occurs ($\mu_0 H_{DMI}$), $D$ can be estimated using the expression:[24]

$$D = \mu_0 H_{DMI} M_s \sqrt{A_{ex}/K_{eff}} \tag{S1}$$

The parameters used in Eq. S1 have been previously assigned in the main text. From the $v(\mu_0 H_x)$ curves shown in Supp. Fig. 1(c), a $D$ of -0.14 mJ/m$^2$ can be extracted for the [Pt/Co(0.35 nm)]$_2$ sample. The left-handedness of the iDMI in this sample (denoted by the negative sign of $D$) is determined from by comparing the responses of ↑ to ↓ and ↓ to ↑ domain walls to the direction in which $\mu_0 H_x$ is applied. While there is substantial debate as to the appropriateness of using $\mu_0 H_x$-induced domain growth asymmetry measurements to determine $D$, previous reports have demonstrated qualitative agreement between $D$ values rendered by the aforementioned technique and other approaches, such as Brillouin light scattering (BLS) spectroscopy.[35,68,69]

**Supplemental Figures:**

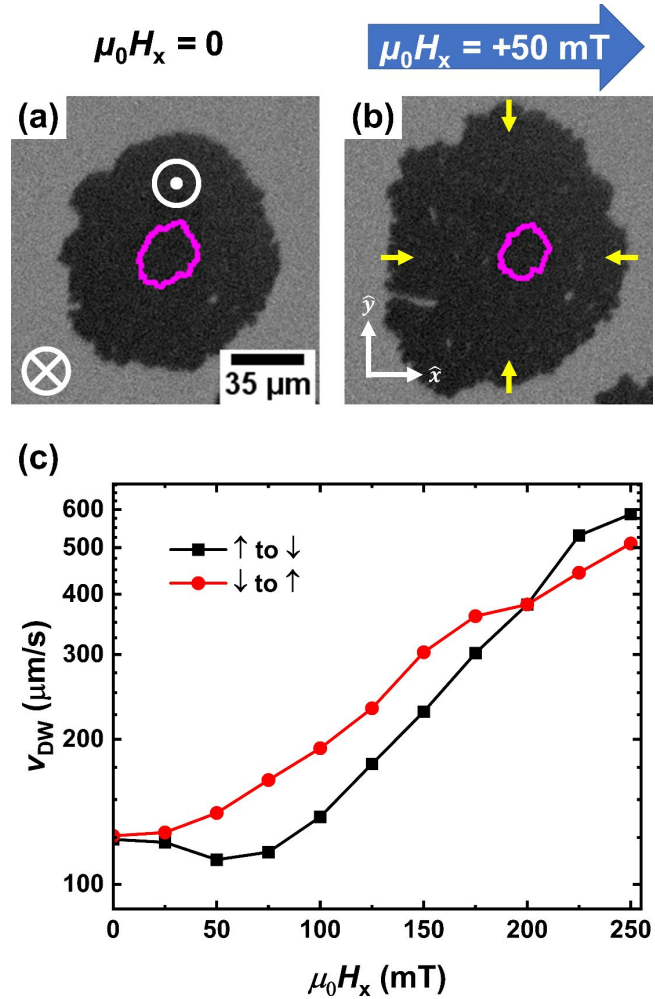

Supplemental Figure 1: Polar MOKE images of domain growth of a $+M_z$ domain in a [Pt/Co(0.35 nm)]$_2$ sample after several out-of-plane field pulses were applied in in-plane fields $\mu_0 H_x$ of **(a)** 0 mT, and **(b)** +50 mT. The magenta outlines indicate the initial extent of the nucleated domain. The yellow arrows in **(b)** schematically depict the left-handed Neel-type domain walls promoted by the Pt-Co interface. **(c)** The expansion velocity of ↑ to ↓ and ↓ to ↑ domains as a function of $\mu_0 H_x$. When collecting data to generate (c), the perpendicular field pulse strength was ~ 0.8 mT, and the pulse duration was 4 ms.

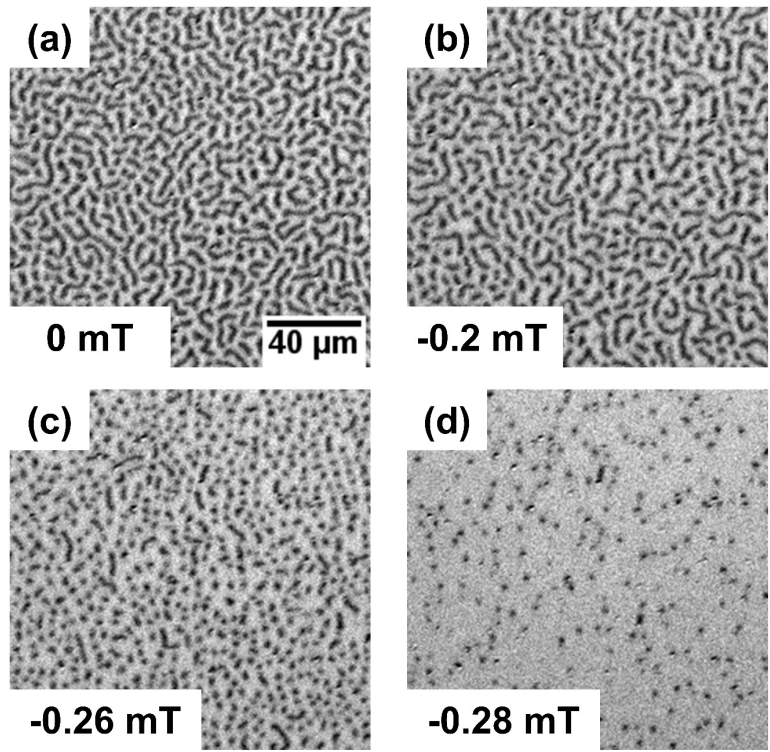

Supplemental Figure 2: Polar MOKE images of the [Pt/Co(0.26 nm)]$_2$ sample, collected as the perpendicular magnetic field strength $\mu_0 H_z$ was quasi-statically stepped from $\mu_0 H_z = 0$ towards negative saturation. When imaging, the sample temperature was approximately 315 K.

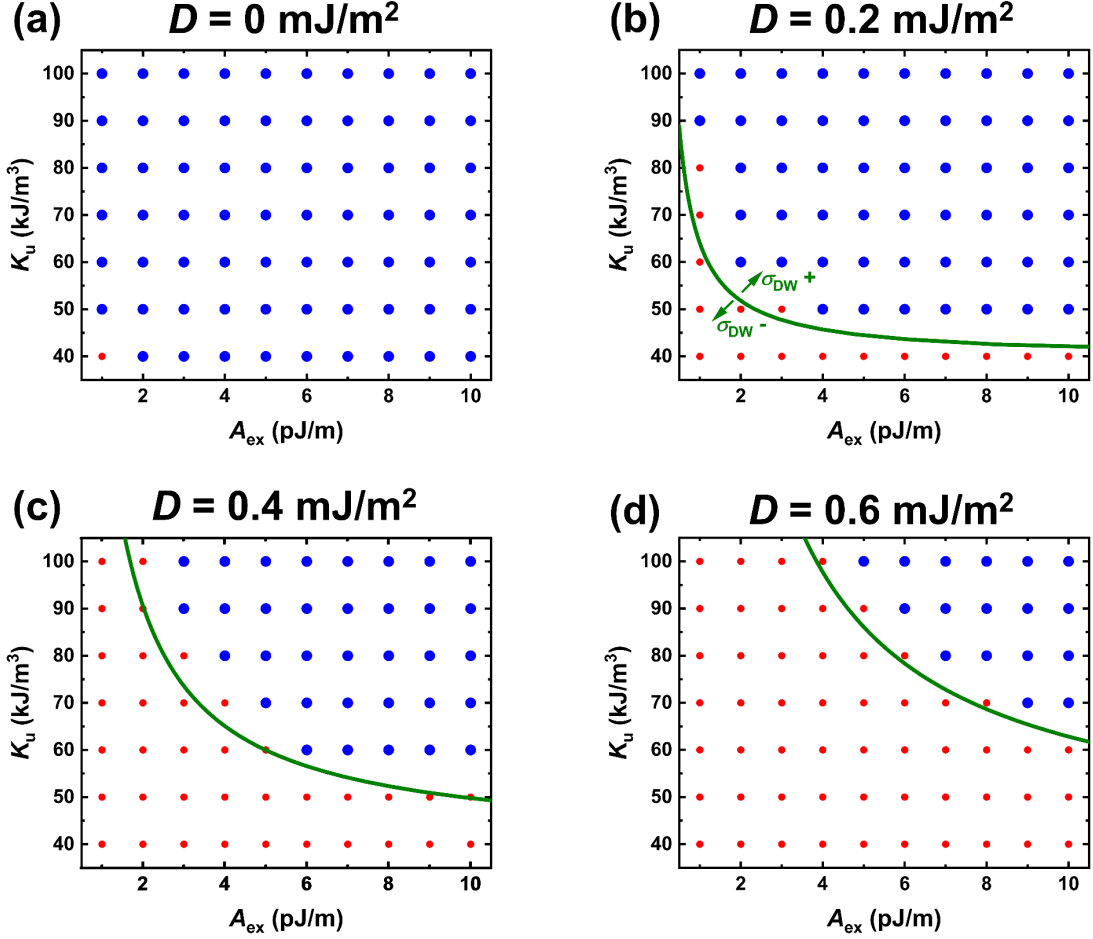

Supplemental Figure 2: $\Delta\varepsilon$ and $\sigma$ as a function of $A_{ex}$ and $K_u$ for different values of $D$, calculated in the manner discussed in the main text. For all calculations, we took $M_s$ as 250 kA/m. The blue (red) circles indicate parameter combinations where the micromagnetic simulations resulted in a positive (negative) $\Delta\varepsilon$. The green line demarcates positive and negative $\sigma$ values, as indicated in **(b)**. Note that in **(a)**, the absence of iDMI dictates that the sign of $\sigma$ does not change for any combination of $A_{ex}$ and $K_u$, hence the lack of a $\sigma$ curve.

**Supplemental Video Captions:**

Supplemental Video 1: Polar MOKE video of the remanent domain morphology of a [Pt/Co(0.24 nm)/Pt]$_1$ sample observed for 20 s.

Supplemental Video 2: Polar MOKE video showing the evolution of the domain morphology in a [Pt/Co(0.24 nm)/Pt]$_2$ sample as a perpendicular magnetic field was quasi-statically stepped from 0 to – 0.8 mT. The video was collected at a framerate of 16 Hz, averaging every 4 frames together. Note: The Co layers in this sample were grown as a wedge using the methods outlined in the main text.

Supplemental Video 3: Polar MOKE video of the motion of skyrmions in a [Pt/Co(0.24 nm)/Pt]$_2$ sample as a current density of 3.7 x 10$^9$ A/m$^2$ was applied from right to left in an applied magnetic field of -0.5 mT . The video was collected at a framerate of 16 Hz, averaging every 4 frames together.

Supplemental Video 4: Polar MOKE video of the motion of skyrmions in a [Pt/ Co (0.24 nm)]$_2$ sample as a current density of 7.5 x 10$^9$ A/m$^2$ was applied from right to left in an applied magnetic field of -0.6 mT. The video was collected at a framerate of 16 Hz, averaging every 4 frames together.

Supplemental Video 5: Polar MOKE video demonstrating the skyrmion Hall effect in a [Pt/ Co (0.24 nm)]$_2$ sample as a current density of 6.5 x 10$^9$ A/m$^2$ was applied for 1 s from left to right in an applied magnetic field of -0.7 mT. The video was collected at a framerate of 16 Hz, averaging every 16 frames together.